\def\ICRR{$^1$}
\def\BU{$^2$}
\def\BNL{$^3$}
\def\UCI{$^4$}
\def\CSU{$^5$}
\def\GMU{$^6$}
\def\GIFU{$^7$}
\def\UH{$^8$}
\def\KEK{$^9$}
\def\KOBE{$^{10}$}
\def\LANL{$^{11}$}
\def\LSU{$^{12}$}
\def\UMD{$^{13}$}
\def\MIYAGI{$^{14}$}
\def\SUNY{$^{15}$}
\def\NIIGATA{$^{16}$}
\def\OSAKA{$^{17}$}
\def\SEOUL{$^{18}$}
\def\TOHOKU{$^{19}$}
\def\TOKYO{$^{20}$}
\def\TOKAI{$^{21}$}
\def\TIT{$^{22}$}
\def\WARSAW{$^{23}$}
\def\UW{$^{24}$}

\def\Superk{Super--Kamiokande}
\def\peppo{\(p \rightarrow e^+ \pi^0\)}
\def\mpeppo{p \rightarrow e^+ \pi^0}
\def\Cerenkov{Cherenkov}

\def\etal{{\it et al}}

\def\exposure{25.5 kton\(\cdot\)year}
\def\efficiency{detection efficiency} 

\newcommand{\limit}[1]{ \(\tau/B_{\mpeppo} > #1 \times 10^{33}\) years}
\newcommand{\psfigure}[1]{\psfig{file=#1,width=6.0cm}}

\documentstyle[aps,prl,psfig]{revtex}

\begin{document}

\draft
\date{\today}
\title{Search for Proton Decay via \peppo{} in a 
  Large Water \Cerenkov{} Detector}
\maketitle
\begin{center}
\newcounter{foots}

The \Superk{} Collaboration  \\

M.Shiozawa\ICRR{}, B.Viren\SUNY{},
Y.Fukuda\ICRR{}, T.Hayakawa\ICRR{}, E.Ichihara\ICRR{}, K.Inoue\ICRR{},
K.Ishihara\ICRR{}, H.Ishino\ICRR{}, Y.Itow\ICRR{},
T.Kajita\ICRR{}
% ...

\footnotesize \it

\ICRR{}Institute for Cosmic Ray Research, University of Tokyo, Tanashi,
Tokyo 188-8502, Japan\\
\BU{}Department of Physics, Boston University, Boston, MA 02215, USA  \\
\BNL{}Physics Department, Brookhaven National Laboratory, 
Upton, NY 11973, USA \\
\UCI{}Department of Physics and Astronomy, University of California, Irvine
Irvine, CA 92697-4575, USA \\
\CSU{}Department of Physics, California State University, 
Dominguez Hills, Carson, CA 90747, USA\\
\GMU{}Department of Physics, George Mason University, Fairfax, VA 22030, USA \\
\GIFU{}Department of Physics, Gifu University, Gifu, Gifu 501-1193, Japan\\
\UH{}Department of Physics and Astronomy, University of Hawaii, 
Honolulu, HI 96822, USA\\
\KEK{}Institute of Particle and Nuclear Studies, High Energy Accelerator
Research Organization (KEK), Tsukuba, Ibaraki 305-0801, Japan \\
\KOBE{}Department of Physics, Kobe University, Kobe, Hyogo 657-8501, Japan\\
\LANL{}Physics Division, P-23, Los Alamos National Laboratory, 
Los Alamos, NM 87544, USA. \\
\LSU{}Physics Department, Louisiana State University, 
Baton Rouge, LA 70803, USA \\
\UMD{}Department of Physics, University of Maryland, 
College Park, MD 20742, USA \\
\MIYAGI{}Department of Physics, Miyagi University of Education, Sendai,
Miyagi 980-0845, Japan\\
\SUNY{}Department of Physics and Astronomy, State University of New York, 
Stony Brook, NY 11794-3800, USA\\
\NIIGATA{}Department of Physics, Niigata University, 
Niigata, Niigata 950-2181, Japan \\
\OSAKA{}Department of Physics, Osaka University, 
Toyonaka, Osaka 560-0043, Japan\\
\SEOUL{}Department of Physics, Seoul National University, 
Seoul 151-742, Korea\\
\TOHOKU{}Department of Physics, Tohoku University, 
Sendai, Miyagi 980-8578, Japan\\
\TOKYO{}The University of Tokyo, Tokyo 113-0033, Japan \\
\TOKAI{}Department of Physics, Tokai University, Hiratsuka, Kanagawa 259-1292, 
Japan\\
\TIT{}Department of Physics, Tokyo Institute for Technology, Meguro, 
Tokyo 152-8551, Japan \\
\WARSAW{}Institute of Experimental Physics, Warsaw University, 00-681 Warsaw,
Poland \\
\UW{}Department of Physics, University of Washington,    
Seattle, WA 98195-1560, USA    \\

\end{center}


\begin{abstract}
  We have searched for proton decay via \peppo{} using data from
  a \exposure{} exposure of the \Superk{} detector.
  We find
  no candidate events with an expected background induced by atmospheric
  neutrinos of 0.1 events. From these data, 
  we set a lower limit on the partial lifetime of the proton
  \(\tau/B_{\mpeppo}\) to be \(1.6 \times 10^{33}\) years
  at a 90\% confidence level.
\end{abstract}




In the Standard Model, which is the modern paradigm of elementary
particle physics, protons are assumed to be stable~\cite{instanton}.
In Grand Unified Theories (GUTs), however, the
decay of the proton is one of the most dramatic predictions of various
models~\cite{su4xsu4,su5gut,pdk_guts1,pdk_guts2}.
In the past two decades,
several large mass underground detector experiments have looked for 
proton decay but no clear evidence has been reported
~\cite{proton.decay.IMB:Phys.Rev.Lett.:1986,proton.decay.kamioka:Phys.Lett.B:1989,proton.decay.IMB:Phys.Rev.D:1990,proton.decay.frejus:Z.Phys.C:1991,proton.decay.frejus:Nucl.Phys.B:1989}.
In general, GUTs predict many modes of proton decay.
In many models, the \peppo{} mode is dominant and there are
several GUTs which predict a decay rate within the observable range of
\Superk{} (see, for
example~\cite{dimension_six1,dimension_six2,dimension_six3}).
This decay mode has a characteristic event signature, in which
the electromagnetic shower caused by the positron is balanced against
the two showers caused by the gamma rays from the decay of the
\(\pi^0\).
This signature enables us to discriminate the signal events clearly
from atmospheric neutrino background.
In this letter, we report the results of our search for proton decay
via \peppo{} in 414 live-days of data between May 1996 and October
1997, corresponding to an exposure of \exposure{} in \Superk{}.
\par

\Superk{} is a large water \Cerenkov{} detector 
located in the Mozumi
mine at 2700 meters-water-equivalent below
the peak of Mt. Ikenoyama in Kamioka, Gifu prefecture, Japan. The detector
holds 50 ktons of ultra-pure water contained in a cylindrical
stainless steel tank measuring 41.4 m in height and 39.3 m in
diameter. The water is optically separated into three concentric
cylindrical regions. 

The 36.2 m high and 33.8 m in diameter inner detector, 
is viewed by 11146, inward-facing, 50~cm diameter
photomultiplier tubes (PMTs). These PMTs uniformly surround the
region giving a photocathode coverage of 40\%.
They were specially developed to have good single photoelectron (p.e.)
response and have a time resolution of 2.5~ns RMS for 1~p.e.
equivalent signals~\cite{PMT}. The PMT signals are
digitized asynchronously by a custom built data acquisition
system~\cite{AMP,TAC/QAC} which can process two successive
signals, enabling us to detect the electron from the decay of a muon.
The system records the number and arrival times of the photons
collected in each PMT. From these values, along with the 
positions of the PMTs, the events are reconstructed.

The 2.0--2.2 m thick outer detector
completely surrounds the inner
detector. This region is viewed by 1885 outward pointing 20
cm diameter PMTs with 60 cm by 60 cm wavelength shifter
plates~\cite{PMTA}. 
The walls of the outer detector are lined with DuPont Tyvek, a
white reflective material, to increase the number of \Cerenkov{}
photons detected.
The primary function of the outer detector is to
veto cosmic ray muons and to help identify contained events.

The 0.5 m thick middle region (dead space) between the inner and 
outer detectors is uninstrumented and is occupied
by the stainless steel support structure as well as water. The
border with the inner detector is lined with opaque black
plastic and the border with the outer detector, by opaque black
low density polyethylene bonded to the reflective Tyvek. 

The trigger we use in this analysis is issued when 29 or more inner
PMTs produce signals greater than 1/4 p.e. in a 200 ns coincidence
window. This trigger threshold corresponds to the mean number of PMTs
hit by the \Cerenkov{} photons from 5.7 MeV electrons. The trigger rate
ranges between 10 Hz and 12 Hz, of which 2.2 Hz is due to cosmic ray
muons entering the inner detector.

\par

The data sample we use for this analysis consists of events which are fully
contained within the inner detector and is identical to that used for
the atmospheric neutrino analysis. (For details,
see~\cite{SK.Sub-GeV}.)
The essential criteria of this selection are as follows:  (1)~no 
significant outer detector activity, (2)~the total number of p.e.'s
in the inner detector is greater than 200,
(3)~the ratio of the maximum number of p.e.'s in a single PMT to the 
total number of p.e.'s in the event is
less than 0.5, (4)~the time interval from the preceding event is greater
than 100 \(\mu\)sec.  

Essentially 100\% of the cosmic ray muons are eliminated by criterion~(1).
Criterion~(2) corresponds to a lower momentum cut of 22
MeV/c for electrons and 190 MeV/c for muons. 
Criterion~(3) removes spurious electrical noise events.
Criterion~(4) removes
electrons from the decay of stopping cosmic ray muons as well as the noise
events caused by unwanted PMT signals following highly energetic
events (``after pulsing''). 
By applying these criteria, the number of events is reduced from about 
400~million to about 12,000.
In addition all events which follow a previous event within 30
\(\mu\)sec are tagged as an electron from the decay of a muon.

After criteria~(1)--(4) are applied, further reduction is done by scanners 
using an interactive graphic event display
to eliminate most of the few remaining cosmic ray muons or other noise
events.  About 6,000 events are classified as fully contained events.
In this analysis, only events with a fitted vertex 
inside of the fiducial volume are used.
This fiducial volume is defined as
all points which are more than 2 m from the inner detector wall.
This volume cut removes any remaining entering cosmic ray muon
events and assures that the performance of the reconstruction
algorithms is uniform throughout the fiducial volume.
We observe 3468 fully contained events in the fiducial volume.
The inefficiency to recover \peppo candidates due to the 
criteria~(1)--(4) and scanning is estimated to be less than 0.1\%.
\par


All measurement of physical quantities of an event such as vertex position, the
number of \Cerenkov{} rings, momentum, particle type and the number of
decay electrons, is automatically performed~\cite{SK.Sub-GeV} by
reconstruction algorithms.
The vertex position is estimated by finding
the position at which the timing residual ((photon arrival
time)-(time of flight)) distribution is most peaked.  
Using a Monte Carlo (MC) simulated data sample, the vertex
resolution for \peppo{} events is estimated to be 18 cm.
To find the rings in an event, the charge, viewed from the vertex as a
function of \{\(\theta,\phi\)\}, is Hough
transformed~\cite{Hough-space}. The resulting space is searched for
peaks, giving the ring centers.
%
%
With \peppo MC, \(44 \%\) of the simulated events 
passing proton decay selection criteria
(described below) are classified as 3--ring events
and \(56 \%\) are classified as 2--ring events.  
The 2--ring classification is primarily for events with one
of the two \(\gamma\) rings taking only a small fraction of the
\(\pi^0\)'s energy or overlapping too much with other rings.
%

The particle identification (PID) classifies a particle as a
showering particle (\(e^\pm, \gamma\)) or a nonshowering particle
(\(\mu^\pm, \pi^\pm\)), using the photon distribution of its \Cerenkov{} ring.
For single ring
events, the particle misidentification probability is 
estimated to be less
than 1.0\% using the atmospheric neutrino MC.  This is
confirmed with stopping cosmic ray muons and their associated decay
electrons.  The PID performance was also checked using a 1 kton water
\Cerenkov{} detector 
with \(e\) and \(\mu\) beams from the 12~GeV proton
synchrotron at KEK~\cite{beamtest}.
However, the misidentification
probability differs between single ring and multi ring events due to 
overlapping rings. 
Using a \peppo{} MC sample this misidentification is estimated to be 2\%.

The momentum reconstruction is
important because an appropriate momentum cut will reject
atmospheric neutrino background but accept proton decay events.  
The momentum is estimated from
the total sum of p.e.'s detected within a 70\(^\circ\)
half opening angle from the reconstructed ring direction.  The number
of p.e.'s collected in each PMT is corrected for light attenuation in
water, PMT angular acceptance and PMT coverage. 
In the momentum reconstruction, we assume that the particle is an
electron for showering particles and a muon for nonshowering
particles.
For single ring events, the reconstructed momentum resolution is
estimated to be \(\pm(2.5/\sqrt{E(\rm{GeV})}+0.5)\%\) for electrons
and \(\pm3\%\) for muons, respectively.
For multi-ring events, the fraction of p.e.'s in each PMT due to each
ring is determined using the expected p.e. distribution.  
Then, the momentum for each ring is determined by the same method used 
for single ring events.
The reconstructed momentum resolution is \(\pm10\%\) for each ring in
the \peppo{} events.  

The energy scale 
stability is checked by the reconstructed
mean energy of decay electrons from stopping cosmic ray muons.  It
varies within \(\pm0.5\)\% over the exposure period.  
The absolute energy scale was checked with
many calibration sources such as electrons from a linear 
accelerator~\cite{SK.LINAC},
decay electrons from stopping cosmic ray muons, stopping cosmic ray 
muons themselves, and the reconstructed mass of \(\pi^0\) events
observed in atmospheric neutrino interactions.  
From comparisons of these sources
and MC simulation, the absolute calibration error is estimated to be smaller
than \(\pm2.5\%\).  

Finally, the efficiency for detection of decay electrons is estimated
to be 80\% for \(\mu^+\) and 63\% for \(\mu^-\) by a Monte Carlo
study.
The difference in these efficiencies is due to \(\mu^-\) capture on
\(^{16}\)O.
This efficiency was confirmed to an accuracy of 1.5\% using stopping
cosmic ray muons.\par

The main sources of background for this analysis are atmospheric
neutrino interactions which could mimic a \peppo{} event.
To estimate the number of
background events, we have developed a detailed MC simulation of
atmospheric neutrino interactions, meson propagation in the \(^{16}\)O nucleus,
and propagation of secondary particles, as well as \Cerenkov{} photons, in
the detector water~\cite{SK.Sub-GeV}.
We use the atmospheric neutrino flux of
Honda \etal.~\cite{honda}.
For neutrino interactions in the detector, the following types of
interaction are simulated: quasi elastic scattering, single--\(\pi\)
production, multi--\(\pi\) production, and coherent single--\(\pi\)
production for both charged current (CC) and neutral current (NC).
For the \peppo{} mode, CC \(\pi\) production is the most
important background because it could produce an \(e^\pm\) accompanied by
a \(\pi^0\).  We use Rein-Sehgal's model to simulate single--\(\pi\)
production~\cite{Rein-Sehgal}.  The pion cross-sections in
\(^{16}\)O are calculated with the model by Oset \etal.~\cite{Oset}.
Propagation of produced particles and \Cerenkov{} light
in water is simulated 
with a GEANT~\cite{ref:GEANT} based custom detector simulator.
Propagation of charged  pions in the detector water is simulated by a custom
simulator~\cite{ref:MC} for less than 500 MeV/\(c\) pions and by 
the CALOR~\cite{ref:CALOR} simulator for more than 500 MeV/\(c\) pions.  
%
%
For the \peppo{} MC, the same simulator is used.  In this, as well as
the atmospheric neutrino MC, the Fermi motion of protons, 
the nucleon binding energy, and pion interactions in \(^{16}\)O are
considered.

The observed atmospheric neutrino flavor ratio (\(\nu_\mu/\nu_e\))
in \Superk{}
is significantly smaller than the expected value~\cite{SK.Sub-GeV}.
For comparison of data and atmospheric neutrino MC, the neutrino
MC sample is normalized to the number
of observed atmospheric neutrino
events at \Superk{} in the following manner.
The number of \(\nu_e\) (\(\nu_\mu\)) CC
events is normalized by the ratio of the number of single ring
events with a showering (nonshowering) PID in the data to the number of 
single ring events with a showering (nonshowering) PID in 
the atmospheric neutrino MC.
For NC events, the same normalization factor as that of the \(\nu_e\)
CC events is used.

\par

To extract the \peppo{} signal from the event sample, 
these selection criteria are
defined: 
(A)~6800 p.e. \(<\) total p.e. \(<\) 9500 p.e., 
(B)~the number of rings is 2 or 3, 
(C)~all rings have a showering PID, 
(D)~85 MeV/\(c^2 < \pi^0\) invariant mass \(<\) 185 MeV/\(c^2\), 
(E)~no decay electron, 
(F)~800 MeV/\(c^2\) \(<\) total invariant mass \(<\) 1050 MeV/\(c^2\)
and total momentum \(<\) 250 MeV/\(c\).  
Criterion (A) roughly corresponds to a total energy of 800 MeV to 1100
MeV.
Criterion~(C) selects \(e^\pm\) and \(\gamma\).
Criterion~(D) only applies to 3-ring events. Here, at least one pair
of rings must give a reconstructed invariant mass which is consistent
with the estimated \(\pi^0\) mass resolution of
\(135\pm35\)MeV/\(c^2\).
Criterion~(E) is required since the desired \(e^+\) and \(\pi^0\)
particles produce no decay electrons.  
In criterion (F), the total momentum is defined as
\(P_{tot} = |\sum_{i}^{all~rings} \vec p_i|\) where \(\vec p_i\)
is reconstructed momentum vector of \(i\)-th ring.  
The total invariant mass
is defined as \(M_{tot} = \sqrt{E_{tot}^2 - P_{tot}^2}\) where
total energy \(E_{tot} = \sum_{i}^{all~rings} |\vec p_i|\).
Criterion~(F) checks that the total invariant mass and total
momentum correspond to the mass and momentum of the source proton,
respectively. 


\begin{figure}[thp]
  \centerline{\psfigure{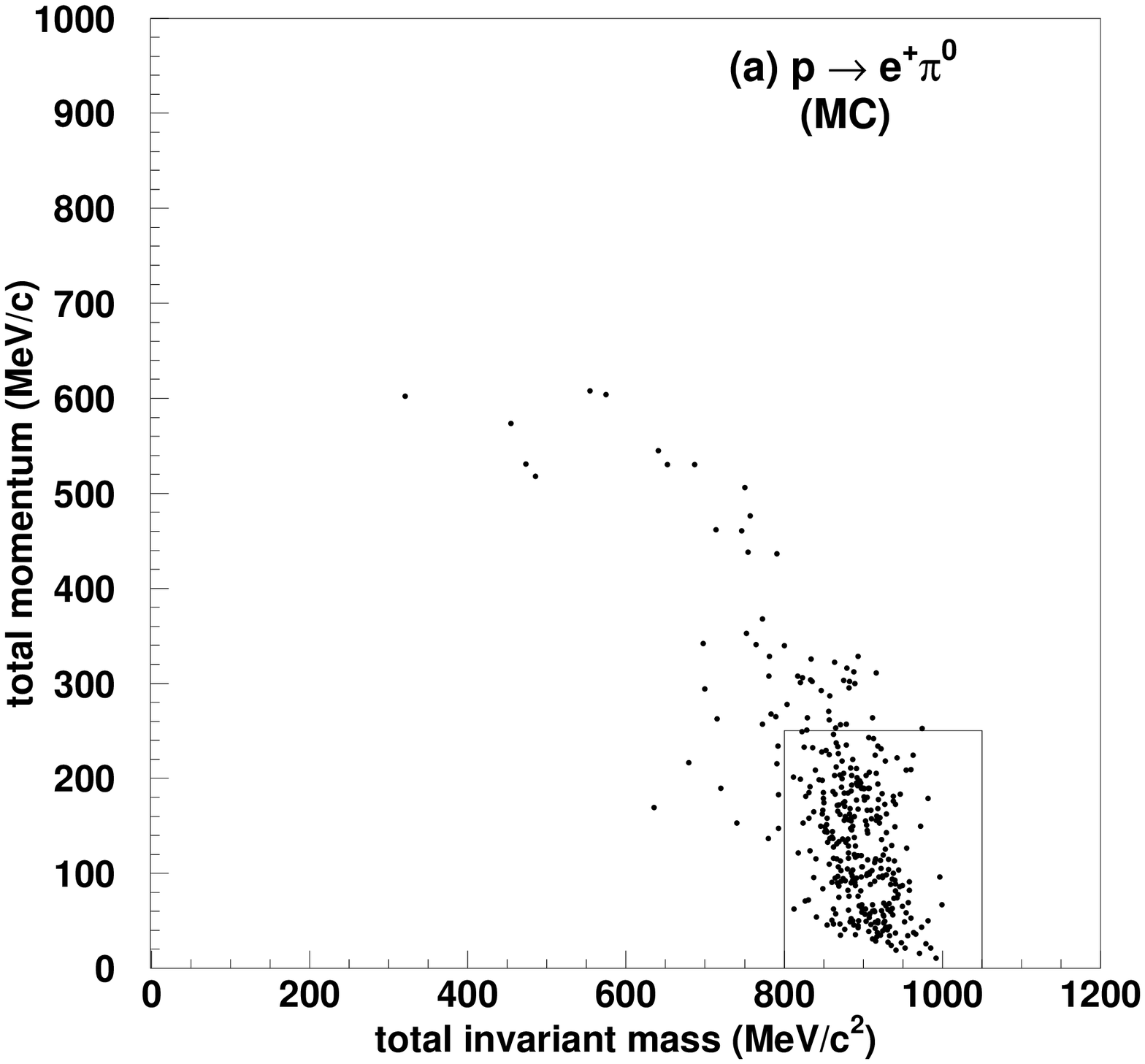}}
  \centerline{\psfigure{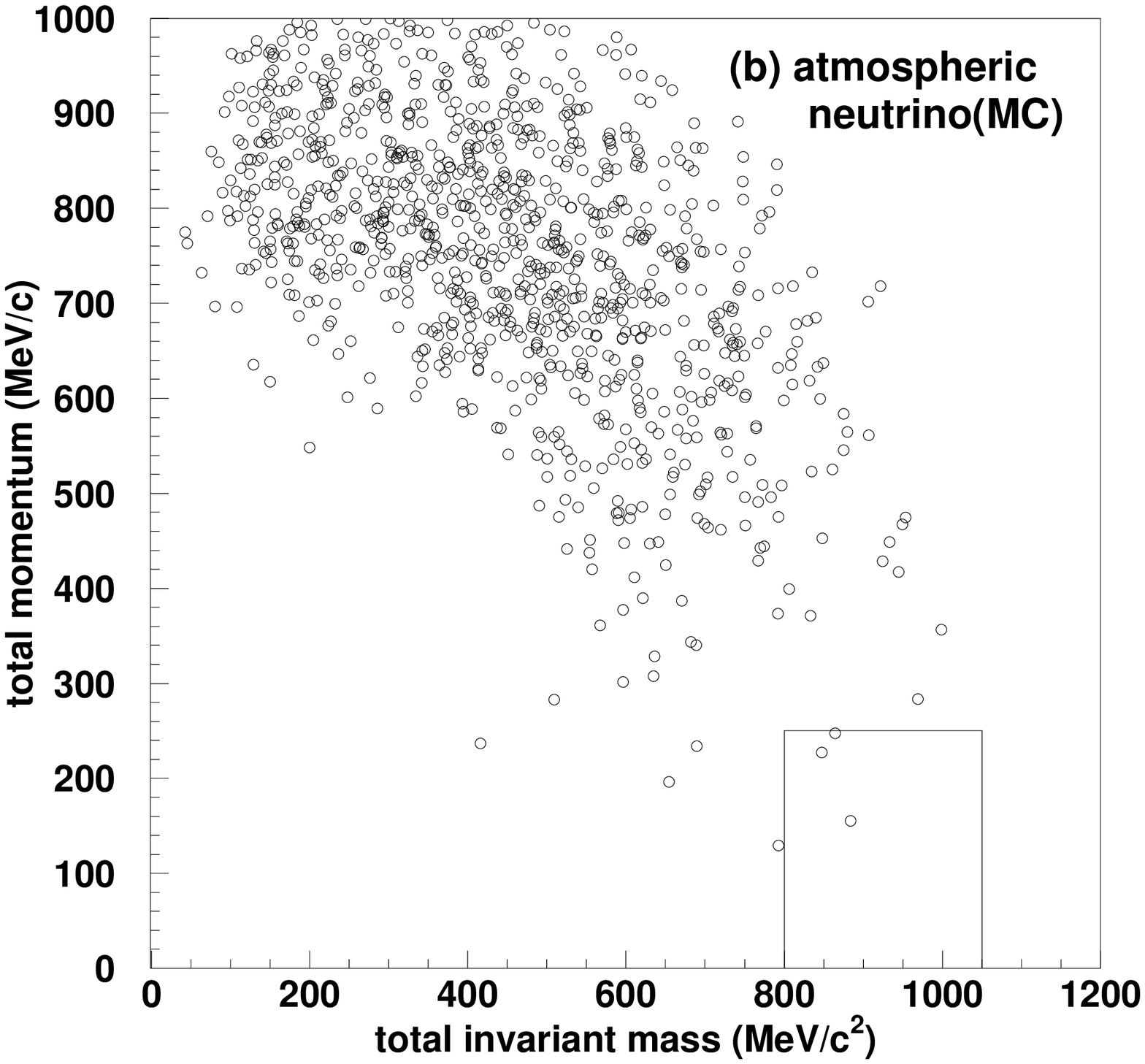}}
  \centerline{\psfigure{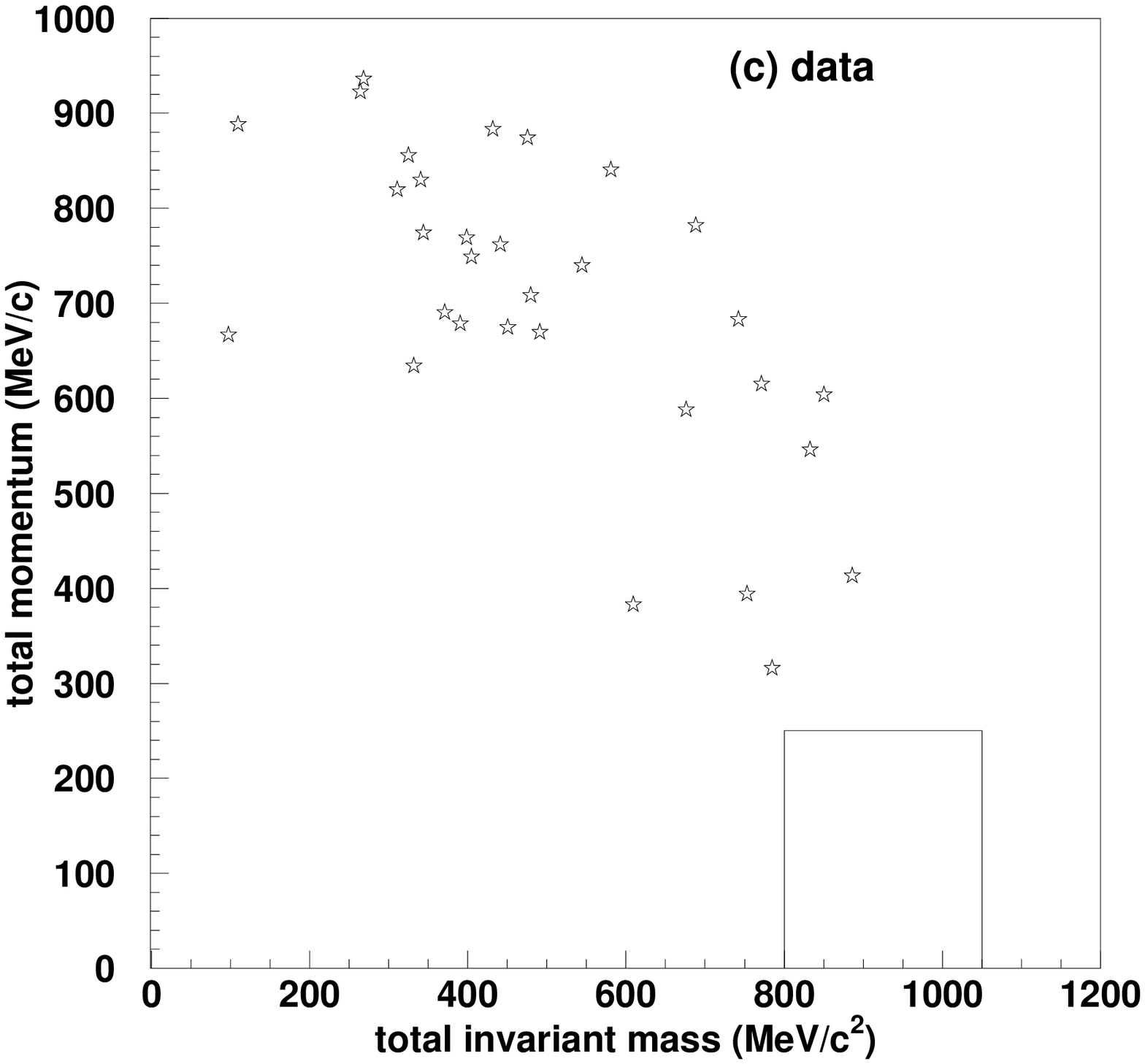}}
  \caption{The total invariant mass and total momentum distributions 
    after criteria
    (A)--(E) (see text) for 3 samples:
    (a) \peppo{} Monte Carlo, (b) atmospheric neutrino Monte Carlo 
    corresponding to 900 kton\(\cdot\)year,
    (c) data corresponding to \exposure{}.  The boxed region in each
    figure shows the criterion (F) for the \peppo{} signal.}
  \label{fig:mass-momentum}
\end{figure}


Figure~\ref{fig:mass-momentum}-(a) shows total invariant mass and total
momentum distributions for the \peppo{} MC sample after criteria
(A)--(E).  
The boxed
region in the figure shows the criterion (F).  
From this sample, the \efficiency{} of \peppo{} events is estimated to
be 44\%.  
The absorption, charge exchange and scattering of \(\pi^0\)'s in the
\(^{16}\)O nucleus are the dominant contribution to the detection
inefficiency.
To estimate the background from atmospheric neutrino interactions, we
generate a MC sample of 900 kton\(\cdot\)year.
By applying the proton decay selection criteria to this sample, we
estimate the number of background events in the signal region to be
\(0.1\) event in \exposure{}.
Figure~\ref{fig:mass-momentum}-(b) shows the total invariant mass and
total momentum distributions of the neutrino MC sample.
\par


\begin{figure}[p]
  \centerline{\psfigure{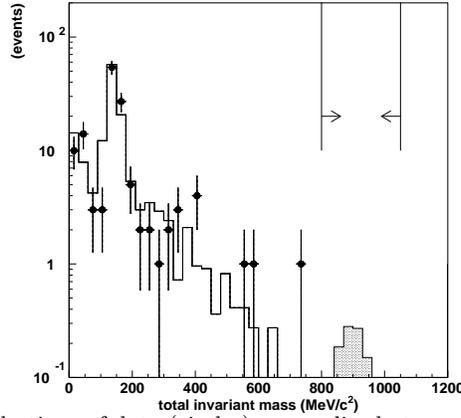}}
  \caption{The total invariant mass distributions of data (circles), 
    normalized atmospheric neutrino Monte Carlo corresponding 
    to 10 years (unshaded histogram),
    and \peppo{} Monte Carlo normalized to one event
    (shaded histogram) which satisfy the criteria
    (B)--(E) (see text) and have a total reconstructed 
    momentum \(<\) 250 MeV/\(c\).}
  \label{fig:mass-MCdata}
\end{figure}



\begin{figure}[p]
  \centerline{\psfigure{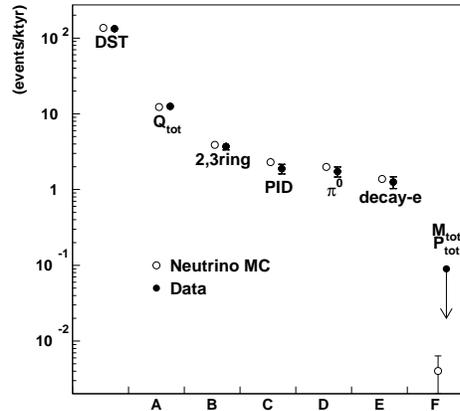}}
  \caption{The event rate after each proton decay selection criterion 
    (see text) for
    data (filled circles) and atmospheric neutrino Monte Carlo (empty
    circles).  There is no event in the data after criterion F and
    only the 90\% C.L. upper limit is shown in the last bin.}
  \label{fig:reduction}
\end{figure}


Finally we apply the same criteria to the data
to search for the \peppo{} signal.  No events survive all
criteria as shown in Figure~\ref{fig:mass-momentum}-(c).  
Figure~\ref{fig:mass-MCdata} compares the
reconstructed total mass for atmospheric neutrino MC, 
the \peppo{} MC, and the data events which satisfy the 
criteria (B)--(E) and
have a total reconstructed momentum \(<\) 250 MeV/\(c\).  
For this comparison criterion (A) is omitted to provide 
enough statistics.
The mass distribution
of data is well reproduced by the neutrino MC.
Figure~\ref{fig:reduction}
shows the event rate after applying each of criteria (A) through
(F) for the data and atmospheric neutrino MC events.  The data are well
represented by the neutrino MC.
\par

As an overall consistency check, a simpler analysis is done. This
analysis, while free of the systematic errors associated with tight, 
ring--fitting--based momentum and invariant mass cuts, allows a
non-negligible background to pass its cuts.
The selection criteria are as follows: (1)~Visible energy is within
200 MeV of the proton rest mass energy. (2)~The light anisotropy
is less than 30\%.  (3)~No electrons from muon decays follow the primary
products. 
The visible energy of an event is the total energy assuming all
\Cerenkov{} light is from electromagnetic showers.
The light anisotropy of an event is a rough measure of the total 
momentum imbalance and is defined as the
magnitude of the normalized vector sum of the unit directions from the
vertex to each PMT, weighted by the charge and corrected for water
attenuation and PMT acceptance.  
For a typical single ring event the light anisotropy is \(\sim\)75\%.
This simple analysis finds 4 events with a background of 3.5, which is
consistent with the 0 candidates and 0.1 background of the primary
analysis.

%
%

From these results, we conclude we do not find any evidence for
proton decay via the mode \peppo{}.  Therefore
we set a lower limit on the partial decay lifetime.
The quantities of this calculation are a \efficiency{} of
44\%, 0 candidate events out of \exposure{} data, 
and 3 background candidates out of 900 kton\(\cdot\)year 
of simulated background MC.
In addition, the uncertainties associated with these
quantities are included in the limit calculation by employing a
Bayesian method~\cite{bayes_limit}, (for details, see \cite{sknote}).
In this method, the prior probability density functions (priors) for the
exposure and \efficiency{} are taken as Gaussian distributions, truncated
to disallow unphysical regions.
The background prior is taken to be a convolution of 
Poisson and Gaussian distributions in order to account for both the
statistical uncertainty of a finite background MC sample size and the
systematic uncertainty in the atmospheric neutrino fluxes and
cross sections used in the background MC sample.
Finally, the prior for the decay rate is taken to be uniform.
This corresponds to the uniform prior implicitly used in simple Poisson
limits~\cite{pdg_limit}.  The resulting limit on the partial lifetime
for \peppo{} is found to be, \limit{1.6} at a 90\% CL.


In calculating the limit, the parameter with the dominant uncertainty is the
\efficiency{}. This uncertainty is primarily due to imperfectly
known pion--nucleon cross sections in \(^{16}\)O nuclei and is estimated by
comparing with another detailed model (based on~\cite{bertini.Phys.C:1972}). 
This uncertainty is estimated
to be 15\%.  In addition, systematic differences in the energy scale
for data and MC contributes 1\%, lack of uniformity in the 
detector gain contributes 2\% and
fitting resolution contributes 5\% to the 
uncertainty in the \efficiency{}. The total uncertainty in the
\efficiency{} is then 16\%.  
The statistical uncertainty in the background is ~60\% due to the small
number of MC background events passing the cuts.
Finally, the uncertainty in the exposure is negligible.

In this letter, we have reported the results of a proton decay search
in \Superk{}.  We have no evidence of
the proton decaying via the mode \peppo{} in the \exposure{} data.
We set the most stringent limit on the partial lifetime of the proton
to be \(1.6 \times 10^{33}\) years at a 90\% CL, which should be
compared with the previous experimental results,
\(2.6\times10^{32}\) years~\cite{proton.decay.kamioka:Phys.Lett.B:1989} and 
\(5.5\times10^{32}\) years~\cite{proton.decay.IMB:Phys.Rev.D:1990}.

%
%
%
%
%
%
%
%

We gratefully acknowledge the cooperation of the Kamioka Mining and Smelting
Company.  The \Superk{} experiment was built from, and has been operated with,
funding by the 
Japanese Ministry of Education, Science, Sports and Culture, and the United
States Department of Energy.


%
%


\end{document}